\documentclass{webofc}
\usepackage[varg]{txfonts}   
\usepackage{natbib}
\usepackage{subfig}
\usepackage{lineno}
\usepackage[utf8]{inputenc}
\usepackage{float}
\begin{document}
\title{A Computing and Detector Simulation Framework for the HIBEAM/NNBAR Experimental Program at the ESS}
%
%
\author{\firstname{First author} \lastname{First author}\inst{1,3}\fnsep\thanks{\email{Mail address for first
    author}} \and
        \firstname{Second author} \lastname{Second author}\inst{2}\fnsep\thanks{\email{Mail address for second
             author if necessary}} \and
        \firstname{Third author} \lastname{Third author}\inst{3}\fnsep\thanks{\email{Mail address for last
             author if necessary}}
}

\institute{Insert the first address here 
\and
           the second here 
\and
           Last address
          }

\author{
 \firstname{Joshua} \lastname{Barrow}\inst{10,11} \and
 \firstname{Gustaaf} \lastname{Brooijmans}\inst{2} \and
 \firstname{José Ignacio Marquez} \lastname{Damian}\inst{3} \and
 \firstname{Douglas} \lastname{DiJulio}\inst{3} \and
 \firstname{Katherine} \lastname{Dunne}\inst{4} \and   
 \firstname{Elena} \lastname{Golubeva}\inst{5} \and
\firstname{Yuri} \lastname{Kamyshkov}\inst{1} \and
\firstname{Thomas} \lastname{Kittelmann}\inst{3} \and
\firstname{Esben} \lastname{Klinkby}\inst{8} \and
\firstname{Zsófi} \lastname{Kókai}\inst{3} \and  
\firstname{Jan} \lastname{Makkinje}\inst{2} \and   
\firstname{Bernhard} \lastname{Meirose}\inst{4,6}\fnsep\thanks{\email{bernhard.meirose@fysik.su.se.}} \and
\firstname{David} \lastname{Milstead}\inst{4} \and
\firstname{André} \lastname{Nepomuceno}\inst{7} \and     
\firstname{Anders} \lastname{Oskarsson}\inst{6} \and 
\firstname{Kemal} \lastname{Ramic}\inst{3} \and 
\firstname{Nicola} \lastname{Rizzi}\inst{8} \and 
\firstname{Valentina} \lastname{Santoro}\inst{3} \and   
\firstname{Samuel} \lastname{Silverstein}\inst{4} \and
\firstname{Alan} \lastname{Takibayev}\inst{3} \and
\firstname{Richard} \lastname{Wagner}\inst{9} \and
 \firstname{Sze-Chun} \lastname{Yiu}\inst{4} \and  
\firstname{Luca} \lastname{Zanini}\inst{3} \and
\firstname{Oliver} \lastname{Zimmer}\inst{9}
       }
\institute{Department of Physics and Astronomy, The University of Tennessee, Knoxville, TN 37996, USA \and  
Department of Physics, Columbia University, New York, NY 10027, USA
 \and  
 European Spallation Source ERIC, 225 92, Lund, Sweden\and  
            Department of Physics, Stockholm University, 106 91, Stockholm, Sweden \and 
Institute for Nuclear Research, Russian Academy of Sciences, Prospekt 60-letiya Oktyabrya 7a,
Moscow, 117312, Russia \and 
            Fysiska institutionen, Lunds universitet, 221 00, Lund, Sweden \and  
            Universidade Federal Fluminense, Rio das Ostras, RJ,
Brazil  \and 
DTU Physics, Technical University of Denmark, 2800 Kgs. Lyngby, Denmark \and  
Institut Laue-Langevin, 38042 Grenoble, France \and 
Massachusetts Institute of Technology, Dept. of Physics, Cambridge, MA 02139, USA \and 
School of Physics and Astronomy, Tel Aviv University, Tel Aviv 69978, Israel 
          }

\abstract{The HIBEAM/NNBAR program is a proposed two-stage experiment at the European Spallation Source focusing on searches for baryon number violation via processes in which neutrons convert to antineutrons. This paper outlines the computing and detector simulation framework for the HIBEAM/NNBAR program. The simulation is based on predictions of neutron flux and neutronics together with signal and background generation. A range of diverse simulation packages are incorporated, including Monte Carlo transport codes, neutron ray-tracing simulation packages, and detector simulation software. The common simulation package in which these elements are interfaced together is discussed. Data management plans and triggers are also described.
} 

\maketitle

\section{Introduction}
\label{sec:intro}
The HIBEAM/NNBAR program~\citep{White} for the European Spallation Source (ESS) is a future sequence of experiments to search for beyond Standard Model (BSM) physics appearing as baryon number violation (BNV). The primary goal of the experiment is to look for free neutron--antineutron transformations ($n\rightarrow \bar{n}$)\footnote{An additional physics goal is the search for neutrons converting to sterile neutrons~\citep{White}, which would employ broadly the same simulation framework as described here.}. It is a two-stage program, the first stage being that of HIBEAM, which plans to utilize a fundamental physics beamline, ANNI~\citep{Soldner:2018ycf}, soon after ESS commissioning and without full beam power. A high precision phase would follow with a dedicated high-flux beamline at the already appropriated Large Beam Port. The HIBEAM/NNBAR experiment has an active collaboration~\citep{White}, and is currently focused on research for a Conceptual Design Report (CDR) to be published by the end of 2023 as a part of a EU Horizon2020 infrastructure design program~\citep{Santoro:2020nke}. The HIBEAM/NNBAR program will ultimately give a sensitivity increase of three orders of magnitude compared with previous neutron transformation searches~\citep{White,BaldoCeolin:1989qd,BaldoCeolin:1994jz}. 



The HIBEAM/NNBAR experimental program is particularly interesting from a computational and simulation perspective as it is a cross-disciplinary experiment with a diverse membership. For example, the collaboration encompasses physicists from both the high-energy hadron collider and low-energy nuclear physics experiments, as well as scientists specializing in neutronics and magnetics. As such, synergies between the communities are continuing to foster solutions to technical problems, including in the computing area, focused upon the common goal of discovering physics beyond the Standard Model via the observation of neutron conversions.  

Presently under construction in Lund, Sweden, the ESS is a next-generation high-flux spallation neutron source. The spallation source of the ESS will emit relatively long $2.8\,$ms neutron pulses with an integrated flux which will greatly exceed that of current facilities once commissioned to its full planned operating power of $5\,$MW. The neutron flux on target for the initial HIBEAM stage at an inaugural power of $\sim2\,$MW is expected to be $\sim10^{11}$n/s~\citep{Soldner:2018ycf,White} soon after commissioning, and the ESS is foreseen to achieve a flux around two orders of magnitude larger for the second stage NNBAR experiment by virtue of the purpose-built Large Beamport which encompasses an opening angle near that of three nominal beamports~\citep{White,Matt-thesis}.

This paper is organized as follows. Section~\ref{sec:signal-data} describes the expected rates of signal and background processes. Section~\ref{sec:chain} then gives an overview of the simulation chain, describing the various component programs. Following this, Section~\ref{Sec:neutronics} outlines the neutron focusing planned for the experiment and simulation thereof. Section~\ref{sec:signal} describes the simulation of the signal and background processes. Section~\ref{sec:mcpl} describes the MCPL simulation format~\citep{MCPL} which is used in this work but is not well known in the high energy physics (HEP) community. Section~\ref{sec:detector} outlines the planned detector for the measurement of the signal annihilation process together with its simulation. Section~\ref{sec:trigger} then outlines trigger strategy. Section~\ref{sec:dm} then describes the data management plan for the work. This is followed by a short summary of the paper. Finally, an Appendix is provided which gives an overview of the various component software packages used in the HIBEAM/NNBAR simulation framework.


\section{Signal, background and data taking rates}
\label{sec:signal-data}
The main goal of the HIBEAM/NNBAR program is to observe the striking signature of the annihilation of a free antineutron with a bound nucleon in a target foil. The antineutron would arise from an extranuclear $n\rightarrow\bar{n}$ within a beam of free neutrons in transit. As a search for a rare BNV process, any definitive discovery would require the observation of at least one free neutron oscillating into an antineutron above what are expected to be, following appropriate trigger and analysis selections, tiny backgrounds, as was achieved at the last such experiment~\citep{BaldoCeolin:1994jz}. The annihilation interaction on a carbon nucleus would lead to a multi-pion final state, with an average of around $4$-$5$ primary 
particles released from the nucleus. These will be captured by a large solid-angle detector, where several charged pions and photons (resulting from heavy resonance or neutral pion decays) must be reconstructed within a short timing-coincidence window. For some brief details of the $n\rightarrow\bar{n}$ signal simulation, see Section~\ref{sec:signal} as based on Refs.~\citep{Golubeva:2018mrz,Barrow:2019viz}.

As the HIBEAM/NNBAR program is searching for a rare process, the average data taking trigger rate is mainly driven by natural and neutron beam-generated backgrounds. The target foil material will likely be carbon, a choice driven by its low neutron capture cross section within the expected cold neutron velocity range of the experimental program (averaging to $\sim1000\,$m/s for the HIBEAM/ANNI beamline~\citep{Soldner:2018ycf}). Given the high neutron fluxes reaching the target of the experiments, one still expects around $10^{5}$ and $10^{7}$ photons/s from neutron capture for HIBEAM and NNBAR, respectively. Such rates are quite manageable for a modern trigger system, but it can provide high counting rates within sub-detector elements, and can be a potential source of background to the signal in random 
coincidence with fast neutrons released during the spallation process, along with cosmic rays. Cosmic ray events were the dominant backgrounds for all previous free neutron transformation experiments~\citep{BaldoCeolin:1989qd,BaldoCeolin:1994jz}, alone accounting for the $3\,$Hz trigger rate, with $2.7\,$Hz coming from cosmic ray muons which evaded the veto (an efficiency of $\sim99.5\%$), and $0.3\,$Hz due to neutral cosmic rays. Alongside the signal process, all of these components are key simulation and computation drivers within the HIBEAM/NNBAR program.

\section{Simulation chain}  
\label{sec:chain} 
In order to simulate the final expected $n\rightarrow\bar{n}$ signature in the annihilation detector, a consistent and connected chain of different simulations codes must be implemented.

\begin{figure*}[ht]
    \centering
    \includegraphics[height=4.75cm]{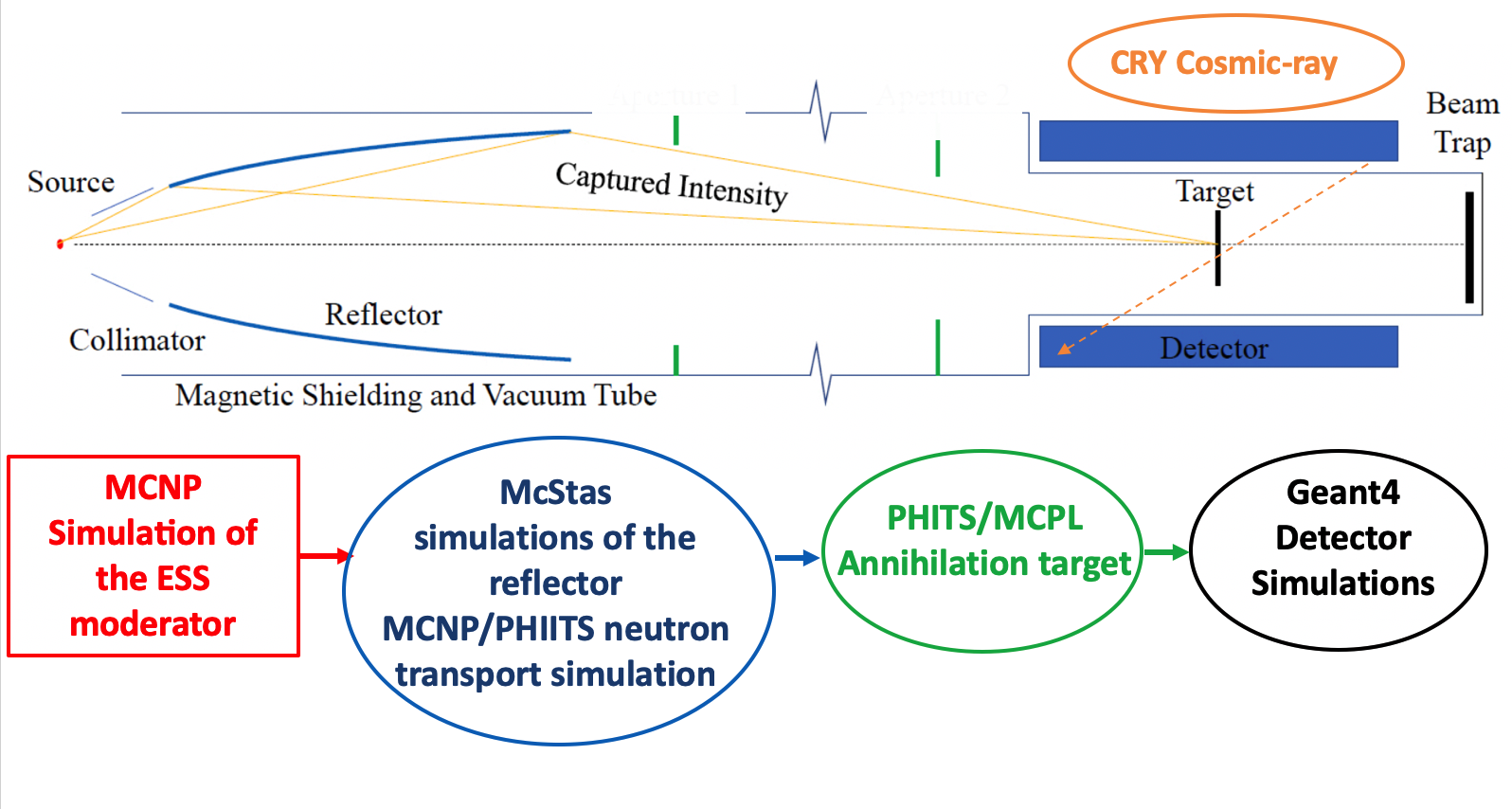}
    \caption{Top: A schematic view of the overall design of the NNBAR experiment, also informative for any BNV searches during the initial HIBEAM stage. Bottom: A HIBEAM/NNBAR simulation flowchart; for the ESS NNBAR Large Beam Port source, the moderator is simulated and optimized with {\fontfamily{qcr}\selectfont MCNP6}. All {\fontfamily{qcr}\selectfont MCNP6} simulation results are passed to {\fontfamily{qcr}\selectfont McStas} with {\fontfamily{qcr}\selectfont MCPL} (see Section \ref{sec:mcpl}). 
    }
    \label{fig:simchain}
\end{figure*}

A graphical
overview of the chain can be seen in the work-flow shown in Fig.~\ref{fig:simchain}, with details given in Sections~\ref{Sec:neutronics} and~\ref{sec:signal}. An overview of the programs is also given in the Appendix.

The chain consists of a number of components, each corresponding to separable aspects of the beamline. The chain begins with the simulation of the ESS' neutron source, following which the reflection and focusing of neutrons must be modelled prior to the simulation of the gravitational propagation of neutrons along 
a vacuum tube. A detector to observe $\bar{n}N$ annihilations\footnote{The letters $A$ and $N$ are used in this work to denote a nucleus and a nucleon, respectively.}, which give a multi-pion final state, must also be simulated, as must this and other interactions on the carbon foil which the detector encloses. Cosmic rays are an external background which require estimation with a dedicated program. Energy and tracking reconstruction software are being developed by the HIBEAM/NNBAR collaboration. The simulation allows the quantification of the experiment’s sensitivity to different assumed values of the $n \rightarrow \bar{n}$ oscillation time.
 
\section{Neutronics and neutronics simulations}\label{Sec:neutronics}
\begin{figure*}[ht]
    \centering
    \includegraphics[height=5.75cm]{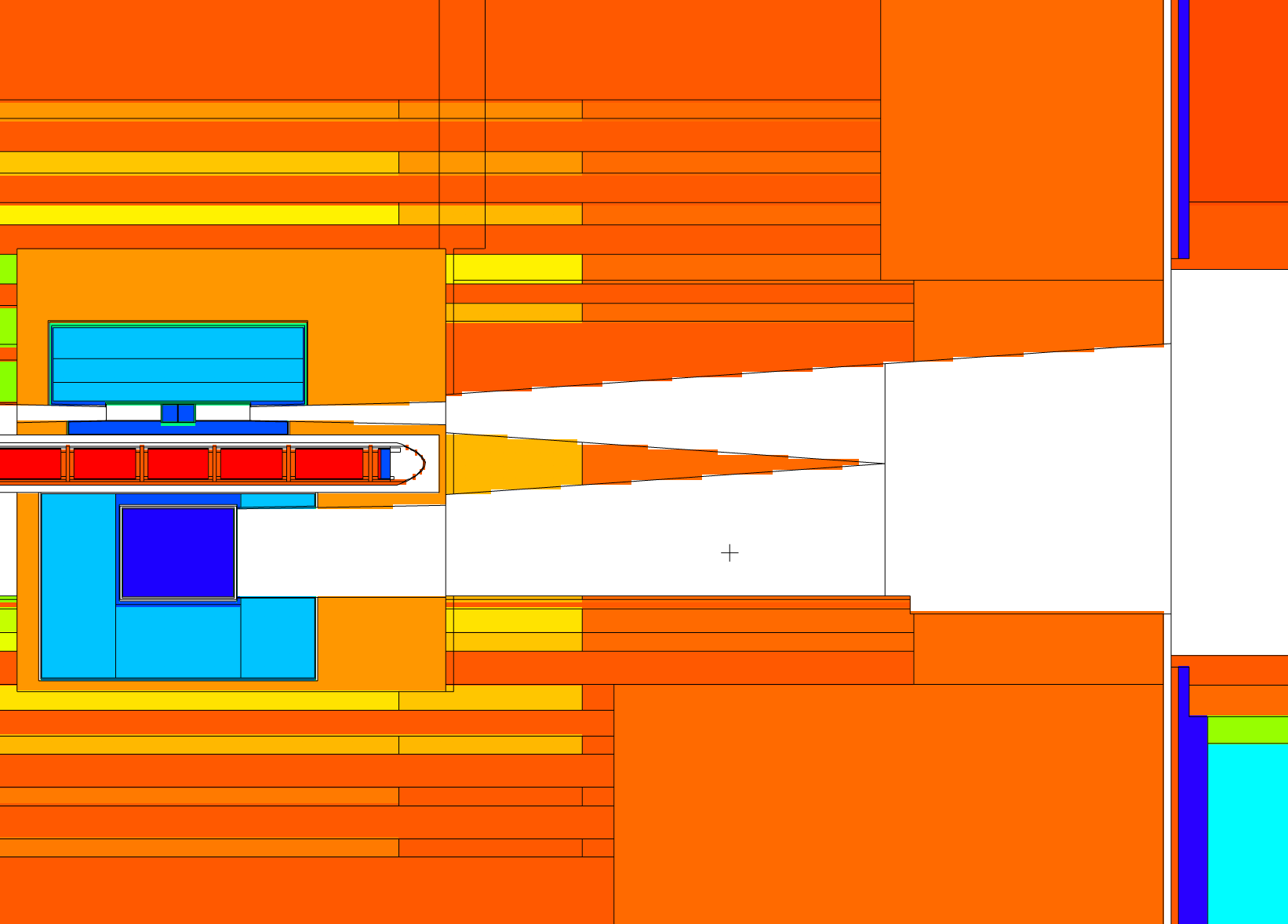}
    \caption{Conceptual design of the neutron source for the NNBAR experiment at the Large Beam Port. A large liquid deuterium moderator (blue) is placed below the tungsten spallation target (red), and is surrounded by a beryllium reflector (light blue). This design is the first concept for the source, and is presently under study as part of the HighNESS project.}
    \label{fig:moderator}
\end{figure*}

Neutron source simulation studies for NNBAR are performed via the radiation transport code {\fontfamily{qcr}\selectfont MCNP6}~\citep{MCNP6} and currently focus on the optimization of the neutron intensity of the source, useful both to the standard mission of the ESS as well as the HIBEAM/NNBAR program. A geometric description of the ESS target and moderator system with the baseline concept for the high-intensity liquid deuterium moderator as implemented within {\fontfamily{qcr}\selectfont MCNP6} can be seen in cross section in Fig.~\ref{fig:moderator}. Several moderator designs and configurations are currently being tested in order to provide the highest possible flux; this is a key output of the Horizon2020 HighNESS program~\citep{Santoro:2020nke}. The source yields from {\fontfamily{qcr}\selectfont MCNP6} are then fed into the neutron ray-tracing code {\fontfamily{qcr}\selectfont McStas}~\citep{mcstas}, thus allowing one to simulate and optimize the performance of various reflector geometries. The interface between the {\fontfamily{qcr}\selectfont MCNP6} simulation and {\fontfamily{qcr}\selectfont McStas} is obtained via {\fontfamily{qcr}\selectfont MCPL}, a binary format which lists particle state (neutron trajectory) information and allows for interchanging of trajectories, particles, or entire events across various Monte Carlo simulation applications; see Section~\ref{sec:mcpl} for more details. The {\fontfamily{qcr}\selectfont McStas} simulation, studying reflector performance, consists of an ellipsoidal-like reflector \cite{Matt-thesis} planned for installation within the first $\sim50\,$m of the beamline, as can be seen in the sketch of Fig.~\ref{fig:simchain}. Along with the {\fontfamily{qcr}\selectfont McStas} simulations, it is also necessary to transport any hypothetically background-inducing higher energy neutrons from the source to the annihilation target location. Such simulations are performed within {\fontfamily{qcr}\selectfont PHITS} and {\fontfamily{qcr}\selectfont MCNP6}, and will soon make use of advanced variance reduction techniques (required due to the long-distance transport of neutrons with low statistics). Once the interactions of the neutron beam and the $n\rightarrow\bar{n}$ signal have been simulated, the corresponding output is interfaced with {\fontfamily{qcr}\selectfont GEANT4}~\citep{Agostinelli:2002hh,Allison:2006ve,Allison:2016lfl}
using {\fontfamily{qcr}\selectfont MCPL}.

\section{Simulation of signal and background processes} \label{sec:signal}
The characteristic semi-spherical topology generated from an $\bar{n}A$ annihilation following extranuclear $n\rightarrow\bar{n}$ can retain up to $\sim1.88\,$GeV of total invariant mass and a rather low total momentum (up to the Fermi momentum) across an average of $4$-$5$ pions expected to be emitted from an event~\citep{Golubeva:2018mrz,Barrow:2019viz}. With the presence of final state interactions (nuclear rescattering and absorption processes taking place during the intranuclear cascade), these can become much fuzzier in reality; see~\citep{Barrow:2019viz} for a detailed discussion. The simulation of this signal is a crucial first step in establishing the detector geometry and material requirements in order that one may reconstruct such an event above any background levels, enabling a definitive discovery. Thus, one must understand the effects of both high and low multiplicities, the tracking efficiency of charged pions, the reconstruction of quickly decaying neutral pions, and the possibility of tracking light, highly ionizing nuclear remnants such as protons or deuterons.

The simulation used by the HIBEAM/NNBAR collaboration for these studies was developed over decades~\citep{Golubeva:1997}, culminating in recent work~\citep{Golubeva:2018mrz,Barrow:2019viz}. Antineutron annihilation data is rather difficult to come by; see~\citep{Bressani:2003pv} for a rather robust review of available data. Only a small portion of all possible $\bar{n}N$ annihilation channels have been measured; therefore, the initial $\bar{n}N$ annihilation branching fractions are taken from experimentally measured channels from available $\bar{n}N$ and (charge-conjugated) $\bar{p}N$ data; for some, connected channels were inferred via isotopic relations, while the rest were taken from a statistical $SU(3)$ model~\citep{Golubeva:2018mrz}. The model for $\bar{N}A$ interaction was systematically tested on available $\bar{p}A$ data, largely from LEAR and other antimatter experiments~\citep{Golubeva:2018mrz}. Thus, one may now simulate extranuclear and intranuclear $\bar{n}A$ annihilations resulting from an $n\rightarrow\bar{n}$ process for a free or bound neutron.

The simulation uses proprietary Monte Carlo code and Fortran libraries. In principle, with foreknowledge of the radial dependence of the annihilation near the surface of the nucleus~\citep{Dover:1996ee}, such events may be simulated for any nucleus, though validation as well as data or model signal comparisons have only been considered for a few species, such as ${}^{12}$C~\citep{Golubeva:2018mrz} (useful for HIBEAM/NNBAR) and ${}^{40}$Ar~\citep{Barrow:2019viz} (useful in DUNE \cite{DUNE}). This is possible given the universal microscopic $\bar{n}A$ annihilation dynamics taken into account within the simulation. The event record can be easily reformatted for consistent integration with a {\fontfamily{qcr}\selectfont GEANT4} simulation using {\fontfamily{qcr}\selectfont MCPL} (see Section~\ref{sec:mcpl}); universal PDG Monte Carlo particle codes have been adopted~\citep{Zyla:2020zbs}. As a cross check, independent analysis comparisons between original signal files and reformatted {\fontfamily{qcr}\selectfont MCPL} event files show identical behavior. Event samples of $\sim$100,000 events or more are available or can be rather quickly produced upon reasonable request from the authors of~\citep{Golubeva:2018mrz,Barrow:2019viz}.

There are different sources of backgrounds which must similarly be simulated, eventually accounting for any of their malicious effects on $n\rightarrow\bar{n}$ signal efficiency. Depending on the background source, the simulation's methodology can vary: 
\begin{itemize}
\item A spallation-generated background, composed of charged and neutral particles, emanates directly from the ESS target. As explained in Section~\ref{Sec:neutronics}, such simulations must make use of Monte Carlo radiation transport codes like {\fontfamily{qcr}\selectfont MCNP6} or {\fontfamily{qcr}\selectfont PHITS} since the particles need to be transported over a long distance; to avoid untenably large amounts of processing, such codes also maintain several variance reductions features which can aid in the propagation of particles in the absence of high statistics.
\item The gamma background due to interaction of the cold neutron beam with the annihilation target (see Section~\ref{sec:detector}). Such backgrounds have been simulated with PHITS but do not require any kind of variance reduction.
\item The cosmic background due to charged and neutral cosmic rays can be simulated with the {\fontfamily{qcr}\selectfont CRY} software library~\citep{CRY} (see Section~\ref{sec:detector}).
\end{itemize} 
\section{Utilizing {\fontfamily{qcr}\selectfont MCPL}} 
\label{sec:mcpl}
The use of {\fontfamily{qcr}\selectfont MCPL} as the common format for interfacing between different simulation codes deserves a special mention since it is not so well-known within the HEP) community. In the case of the HIBEAM/NNBAR experimental program, one of the main advantages {\fontfamily{qcr}\selectfont MCPL} provides is an easy way to interface neutron transport codes' results obtained in {\fontfamily{qcr}\selectfont PHITS}, {\fontfamily{qcr}\selectfont MCNP6}, or neutron optics simulations performed with the neutron ray-tracing code {\fontfamily{qcr}\selectfont McStas}~\citep{mcstas}. It is equally simple to convert proprietary code results to an {\fontfamily{qcr}\selectfont MCPL} format via the use of simple loops and functions, most of which are directly provided within the {\fontfamily{qcr}\selectfont MCPL} documentation. MCPL can, therefore, be used to interface a large variety of codes, including many general purpose Monte Carlo simulation codes commonly used throughout HEP. Though arguably not necessary, having various codes and their output interfaced across a practically universal software tool is both pragmatic and efficient.  

\section{Detector components and simulation} 
\label{sec:detector}
The simulation needs for the HIBEAM/NNBAR program are mostly driven by the reconstruction of the annihilation event in a detector, as well as computing of topological, kinematic, and timing characteristics of any backgrounds for their comparative rejection; neutron optics (for instance, the reflection of neutrons downbeam from reflective ellipsoidal mirrors), as well as magnetic and radiation shielding along the beamline, are also major contributors to the overall computational scheme. Detector design is guided by the need for a high signal efficiency for antineutron detection via a high multiplicity pion reconstruction, along with the ability to maintain a low background yield. 
The main detector components are:

\begin{itemize}
\item An annihilation target, likely a $100\,\mu$m thick carbon disk, with a $1\,$m diameter for the initial HIBEAM stage, and $2\,$m diameter for the full NNBAR second stage.

\item A charged particle tracker, necessary for the determination of pion momenta and the annihilation vertex.

\item A calorimeter for photon and pion energy measurement. The current baseline proposal is a hybrid design, with the innermost part composed of multiple layers of plastic scintillator staves for measuring the energy of soft hadrons, behind which lies a layer of lead glass counters to measure the energy deposited by charged particles via Cherenkov radiation. The novel design is in response to the challenging energy range. It is possible the energy resolution in traditional sampling calorimeters commonly used in high energy physics experiments would be too poor due to statistical fluctuations in energy deposits. A simulation of a sandwich calorimeter with alternating sheets of lead and scintillator with thicknesses optimized for hadrons and photons with energies below 500\,MeV will be used to benchmark the proposed baseline design.

\item  A selective trigger system to gather signal and signal-like background candidate events. 

\item A cosmic ray background veto system.
\end{itemize}
\noindent
In order to simulate annihilation signal events in the detector, along with any background sources, a coherent software chain was developed for the HIBEAM/NNBAR program. Expanding on the flowchart of Figs.~\ref{fig:simchain}:

\begin{itemize}
\item[1.] Signal events are provided by the authors of Ref.~\citep{Golubeva:2018mrz,Barrow:2019viz} and converted into a {\fontfamily{qcr}\selectfont MCPL}-compatible text format~\citep{MCPL}.

\item[2.] The {\fontfamily{qcr}\selectfont MCPL} events are then interfaced with the NNBAR {\fontfamily{qcr}\selectfont GEANT4} detector simulation.

\item[3.] The neutron beam, which is used as the source in {\fontfamily{qcr}\selectfont PHITS}~\citep{phits}, is simulated in {\fontfamily{qcr}\selectfont McStas}. For HIBEAM, the ANNI beam simulation is provided by the authors of Ref.~\citep{Soldner:2018ycf} and interfaced with {\fontfamily{qcr}\selectfont PHITS} via {\fontfamily{qcr}\selectfont MCPL}. It should be noted that this versatility allows for any other Monte Carlo transport code to be used, including {\fontfamily{qcr}\selectfont MCNP6}~\citep{MCNP6}. 

\item[4.] Prompt gamma emission from neutron capture processes with a $^{12}$C target are simulated with {\fontfamily{qcr}\selectfont PHITS} and also interfaced with the {\fontfamily{qcr}\selectfont GEANT4} detector via {\fontfamily{qcr}\selectfont MCPL}. 

\item[5.] Cosmic ray events are interfaced with the detector simulation directly using the {\fontfamily{qcr}\selectfont CRY}~\citep{CRY} library, but externally produced cosmic ray sources may also be used, including those of real measured data.
\end{itemize}



The energy ranges of the physical processes involved in $n\rightarrow\bar{n}$ are far below the typical energy ranges of a high energy collider experiment, but the readout systems are expected to be typical of a high energy experiment. Furthermore, external tracking is planned to be accomplished via a Time Projection Chamber (TPC), while internal tracking is planned to be carried out with silicon strips (though this is still under investigation). Given these, {\fontfamily{qcr}\selectfont GEANT4} appears as natural choice for detector simulation, material inquiry, geometric testing, and optimization. 

\begin{figure*}[ht]
    \centering
    \includegraphics[height=5.15cm]{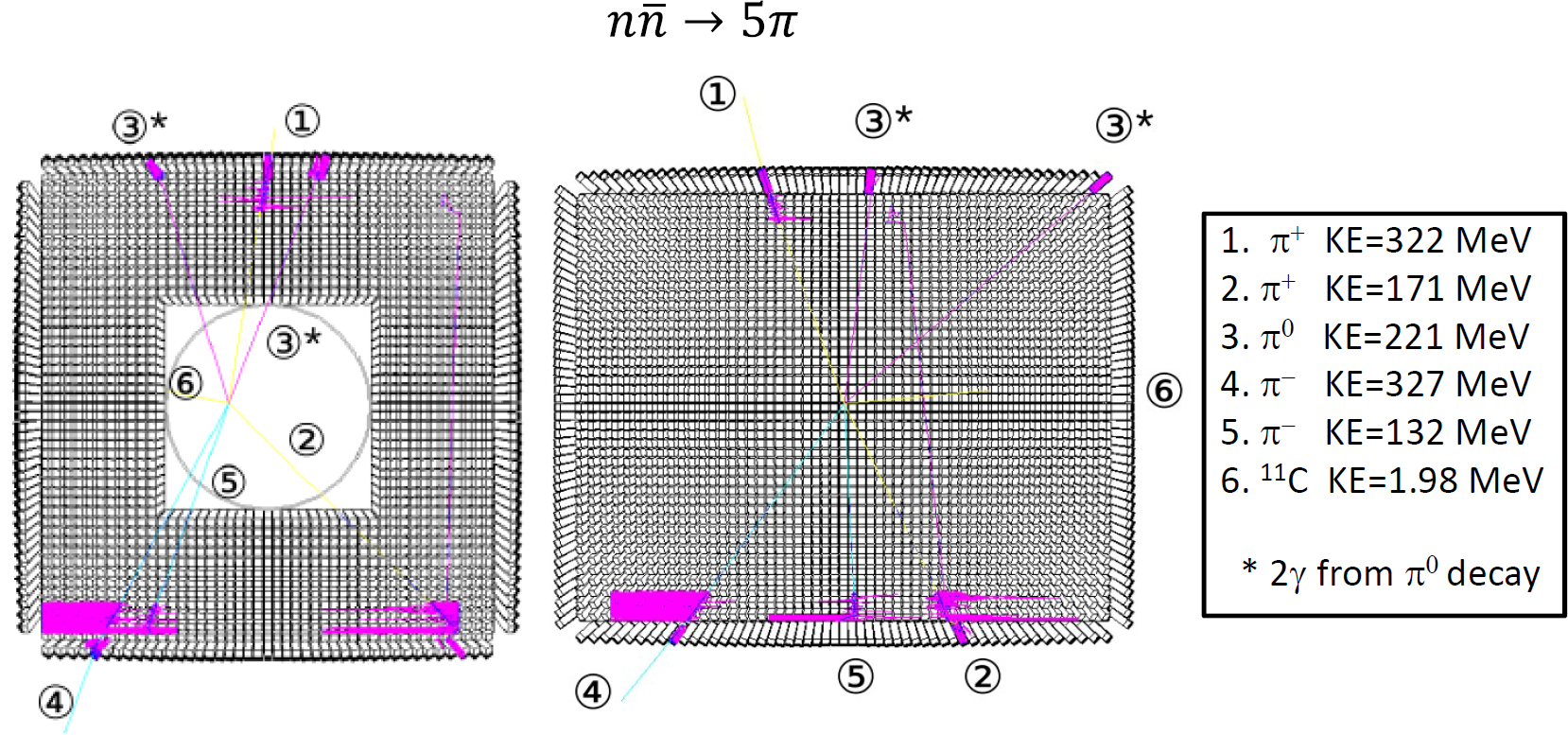}
    \caption{ An event display using {\fontfamily{qcr}\selectfont GEANT4} of a $\bar{n}n$ annihilation process within a $^{12}$C nucleus in a target foil surrounded by an annihilation detector, as modelled with {\fontfamily{qcr}\selectfont GEANT4}. Front (left) and side (centre) views are shown. The truth-level particle identities and kinematic energies of the outgoing annihilation products are given in the box to the right. }
    \label{fig:eventdisplay}
\end{figure*}

Fig.~\ref{fig:eventdisplay} shows an event display of a $\bar{n}{}^{12}$C annihilation signal within {\fontfamily{qcr}\selectfont GEANT4}, as provided by the authors of Ref.~\citep{Golubeva:2018mrz,Barrow:2019viz}. The interaction of annihilation-generated particles with the lead glass blocks of a box detector, shown on the outer edges in purple, are due to Cherenkov light production, wherein a neutral pion is visible through its decay products (two $\gamma$s). A nuclear fragment ($^{11}C$) is also shown in the event, but does not have enough kinetic energy to leave any visible signal within the lead glass. The detector design is merely illustrative - other detector designs are being considered by the collaboration, each attempting to study potential optimizations for signal efficiency, cost, ease of construction and modularity, serviceability and maintenance, as well as long-term replacement of components.

\section{Trigger}\label{sec:trigger}
A two-level trigger is planned for the the second stage full NNBAR experiment. The first level will be hardware based, with self-triggered readout of individual calorimeter and cosmic veto sensors when they detect a particle hit. The readout of each channel will include a global time stamp along with several digitized samples of the shaped analog pulses, allowing both the pulse amplitude and timing to be reconstructed (the latter to ns-level precision).

The second level of the trigger will be implemented off-detector in a CPU farm that will receive the time-stamped readout of the hardware trigger, identify event candidates, and select the corresponding TPC data frames for further event reconstruction and storage. 

 Precise global timing measurements are expected to be one of the main tools for event reconstruction, signal event selection, and background rejection. Important timing applications will include: 
 
\begin{itemize}
\item[1.] Setting short coincidence windows for multiple pion hits with the same origin.
\item[2.] Estimating hit positions in the long scintillator staves from the timing of pulses at opposite ends. 
\item[3.] Hit coincidences to accurately match event data from different subdetectors, for instance, the scintillator and the lead glass calorimeter layers.
\item[4.] Cosmic background rejection using time-of-flight measurements from calorimeter and veto data, to reject particle tracks originating outside the inner detector volume.  
\item[5.] TPC frame selection for matching recorded tracks to calorimeter hits, using the precise calorimeter timing measurements.
\end{itemize}

Pulse shaping and digitization of the calorimeter and veto front-end electronics are foreseen to be implemented with commercial amplifier and ADC devices, along with  programmable logic devices (FPGAs) to provide the digital timing, and self-triggered readout to the off-detector systems. All on-detector components will need to be radiation qualified for the expected particle fluences over the lifetime of the experiment.

By reading out only self-triggered calorimeter and veto channels, the required readout bandwidth for these subdetectors is expected to be relatively modest. This may allow the use of a commercial network switch for connecting the front-end readout with the off-detector processor farm, with PMT samples and timestamp data encoded in compatible network packet formats.

\section{Data Management}\label{sec:dm}
The NNBAR experiment plans to make use of the open-source distributed data management system {\fontfamily{qcr}\selectfont Rucio}~\citep{Rucio}, originally developed for the ATLAS experiment at CERN. {\fontfamily{qcr}\selectfont Rucio} facilitates data storage, management, and processing in a heterogeneous distributed environment. This will be particularly important for the second stage NNBAR experiment, as the scientific data of the experiment is to be shared between the participating institutes\footnote{The general ``white paper"~\citep{White} for the HIBEAM/NNBAR program is signed by authors from 38 institutions, including the ESS, from across 15 countries.}.   
Moreover, there is plenty of experience within the NNBAR collaboration using {\fontfamily{qcr}\selectfont Rucio}, given that several members of NNBAR are either active or former members of ATLAS, including both co-spokespersons.



\section{Summary}  
The HIBEAM/NNBAR experiment is an active experimental program for the European Spallation Source to search for baryon number violation via free neutron--antineutron transformations. The program is currently focused on a series of simulations that will make up a critical part of its CDR that will be delivered by the end of 2023. The full simulation chain is dependent on different components which come together under a consistent framework, which has been described in this paper.

\section{Acknowledgements} 
This work is in part funded by the European Union Framework Programme for Research and Innovation Horizon2020 initiative for the HighNESS project, under grant agreement 951782. The authors also gratefully acknowledge project grant support from Vetenskapsrådet. JB’s work on this project was partially supported by the Visiting Scholars Award Program of the Universities Research Association. Any opinions, findings, and conclusions or recommendations expressed in this material are those of the authors and do not necessarily reflect the views of the Universities Research Association, Inc. JB also wishes to thank the Zuckerman Postdoctoral Scholars Program for their support in finishing this work.

\bibliography{chep20201_bib} 
\newpage
\appendix
\section{Software glossary and overview}\label{sec:software}
Table~\ref{tab:sw} gives an overview of the various software packages used in the HIBEAM/NNBAR simulation framework.
{\small
\begin{table}[H]
\begin{center}
\begin{tabular}{| c | l |l  | c| }
 \hline
 {\bf Software} & {\bf Type of simulation}  & {\bf Use in HIBEAM/NNBAR} & {\bf Ref.}  \\
 \hline
 {\fontfamily{qcr}\selectfont McStas} &  Neutron ray tracing  &  Simulation of neutron reflectors & \citep{mcstas}  \\
  \hline
 {\fontfamily{qcr}\selectfont MCNP6} & Radiation transport in matter & Moderator simulation, &  \citep{MCNP6}   \\
 {\fontfamily{qcr}\selectfont PHITS}  &                               & propagation of neutrons following focusing, &  \citep{phits}  \\
                                     &                                & spallation background propagation, and & \\ 
                                     &                                 & backgrounds from neutron capture & \\    
 \hline
 {\fontfamily{qcr}\selectfont GEANT4} & Radiation transport in matter  & Simulation of annihilation detector for &  \citep{Agostinelli:2002hh,Allison:2006ve,Allison:2016lfl}  \\
                                      &                                & signal and background processes & \\
 \hline
 Signal  & Annihilation Monte Carlo  & Simulation of $\bar{n}{}^{12}$C annihilation  &  \citep{Golubeva:2018mrz,Barrow:2019viz} \\
 generator & model for $\bar{n}N$ & &    \\
 \hline 
{\fontfamily{qcr}\selectfont CRY} & Cosmic ray  generator & Cosmic ray background simulation &  ~\citep{CRY} \\
 \hline 
 {\fontfamily{qcr}\selectfont MCPL} &  Monte Carlo model interface & Interface between {\fontfamily{qcr}\selectfont GEANT4,PHITS,}  & \citep{MCPL}  \\ 
  & & {\fontfamily{qcr}\selectfont McStas,} and {\fontfamily{qcr}\selectfont MCNP6} &   \\
\hline
\end{tabular}
\end{center}
\caption{Description of the various component software packages used in the HIBEAM/NNBAR simulation framework. }
\label{tab:sw}
\end{table}
}

\end{document}